\documentclass[onecolumn,floatfix,superscriptaddress,a4paper,
               showpacs,showkeys,nofootinbib,preprint]{revtex4}

\usepackage{epsfig}
\usepackage{latexsym}
\usepackage{xspace}
\usepackage{hyperref}
\usepackage[latin2]{inputenc}
\usepackage{indentfirst}
\usepackage{enumerate}

\usepackage{amsmath}
\usepackage{amssymb}
\usepackage[english]{babel}
\usepackage{url}

\newcommand{\bs}{\boldsymbol}

\begin{document}

\title{\bf Selfinteracting Particle-Antiparticle System of Bosons}
\author{D. Anchishkin}
\affiliation{
Bogolyubov Institute for Theoretical Physics, 03143 Kyiv, Ukraine}
\affiliation{
Taras Shevchenko National University of Kyiv, 03022 Kyiv, Ukraine}

\author{V. Gnatovskyy}
\affiliation{
Taras Shevchenko National University of Kyiv, 03022 Kyiv, Ukraine}
\author{D. Zhuravel}
\affiliation{
Bogolyubov Institute for Theoretical Physics, 03143 Kyiv, Ukraine}
\author{V. Karpenko}
\affiliation{
Taras Shevchenko National University of Kyiv, 03022 Kyiv, Ukraine}



\begin{abstract}
Thermodynamic properties of a system of interacting boson particles
and antiparticles at finite temperatures are studied within the
framework of the thermodynamically consistent Skyrme-like mean-field
model.
The mean field contains both attractive and repulsive terms.
Self-consistency relations between the mean field and thermodynamic
functions are derived.
We assume a conservation of the isospin density for all temperatures.
It is shown that, independently of the
strength of the attractive mean field, at the critical temperature
$T_{\rm c}$ the system undergoes the phase transition of second
order to the Bose-Einstein condensate, which exists in the
temperature interval $0 \le T \le T_{\rm c}$.
We obtained  that the condensation represents a discontinuity of the
derivative of the heat capacity at  $T = T_{\rm c}$, and condensate occurs
only for the component with a higher particle-number density in the
particle-antiparticle system.
\end{abstract}

\maketitle

\section{Introduction}
\label{sec1}

Knowing the phase structure of the meson systems in the regime of finite
temperatures and isospin densities is crucial for understanding a wide
range of phenomena from nucleus-nucleus collisions to neutron stars and
cosmology.
This field is an essential part of investigations of hot and dense
hadronic matter, which is a subject of active research \cite{bzdak-esum-2020}.
Meanwhile, the meson systems' investigations have their specifics due to a
possibility of the Bose-Einstein condensation of interacting bosonic particles.
The problem of the Bose-Einstein condensation of $\pi$-mesons   
has been studied previously, starting from the pioneer works of
A.B.~Migdal and coworkers (see \cite{anchishkin-mishustin-2019} for
references).
Later this problem was investigated by many authors using different models
and methods.
The formation of classical pion fields in heavy-ion collisions was discussed in
refs.~\cite{anselm-1991,blaizot-1992,bjorken-1992,mishustin-greiner-1993}
and the systems of pions and K-mesons  with a finite isospin
chemical potential have been considered in more recent studies
\cite{son-2001,kogut-2001,toublan-2001,mammarella-2015,carignano-2017,mannarelli-2019}.
First-principles lattice  calculations provide a solid basis for our
knowledge of the finite temperature regime.
Interesting new results concerning dense pion systems have been obtained
recently using lattice methods \cite{brandt-2016,brandt-2017,brandt-2018}.

In the present paper we consider interacting particle-antiparticle
boson system at the conserved isospin density $n_I$ and finite
temperatures.
We name the bosonic particles as ``pions'' just conventionally.
The preference is made because the charged
$\pi$-mesons are the lightest hadrons that couple to the isospin
chemical potential.
On the other hand, the pions are the lightest nuclear boson particles, and
thus, an account for ``temperature creation'' of particle-antiparticle pairs
is a relevant problem based on the quantum-statistical approach.

To account for the interaction between the bosons we introduce a
phenomenological Skyrme-like mean field $U(n)$, which depends only
on the total meson density $n$.
This mean field rather reflects the
presence of other strongly interacting particles in the system, for
instance, $\rho$-mesons and nucleon-antinucleon pairs at low
temperatures or gluons and quark-antiquark pairs at high
temperatures, $T > T_{\rm qgp} \approx 160$~MeV.
Calculations for noninteracting hadron resonance gas show that the
particle densities may reach values $(0.1 - 0.2)$~fm$^{-3}$ at
temperatures $100 - 160$~MeV, which are below the deconfinement
phase transition, see e.g. refs.~\cite{satarov-2009,vovchenko-2017}.

The presented study is  a development of the approach proposed
in ref.~\cite{anchishkin-mishustin-2019}, where the boson system was considered
within the framework of the Grand Canonical Ensemble with zero chemical potential.
Meanwhile, here  we investigate the thermodynamic properties of the meson system
in the Canonical Ensemble, where the canonical variables are the temperature
$T$ and the isospin density $n_I$.
We regard a studied self-interacting many-particle system as a toy model that
can help us understand Bose-Einstein condensation and phase transitions over
a wide range of temperatures and densities.

So, in this work, in the formulation of the Canonical Ensemble, we calculate
the thermodynamic characteristics of a non-ideal hot ``pion'' gas with a
fixed isospin density  $n_I = n_\pi^{(-)} - n_\pi^{(+)} > 0$, where
$n_\pi^{(\mp)}$ are the particle-number densities of the $\pi^{\mp}$ mesons,
respectively.

We hope that our approach, which is physically transparent and clear enough,
will help understand more complex pictures of the phase structure of mesonic
systems arising in quark-meson models, for example, in the
Nambu-Yona-Lasinio model and the lattice calculations.

In Sect.~\ref{sec:mfm-part-antipart} we develop the formalism of the thermodynamic
mean-field model \cite{anch-vovchenko-2015} to describe the boson system of
particles and antiparticles, which will be used in the presented calculations.
In Sect.~\ref{sec:skyrme-param}, we introduce a Skyrme-like parametrization
of the mean field, and after solving the system of self-consistent equations,
we calculate the thermodynamic functions.
In Sect.~\ref{sec:thrmod-properties} we demonstrate the possibility of
Bose condensation when the attractive interaction is ``weak''.
Furthermore, we determine that this is a second-order phase transition.
Our conclusions are summarized in Sect.~\ref{sec:conclusions}.

\section{The mean-field model for the system  \newline
of boson particles and antiparticles}
\label{sec:mfm-part-antipart}

The consideration in this section is based on the thermodynamic mean-field
model, which was introduced in refs.~\cite{anch-1992,anchsu-1995}, and then
developed in ref.~\cite{anch-vovchenko-2015}.
We limit our consideration to the case where at a fixed temperature,
the interacting boson particles and boson antiparticles are in the
dynamical equilibrium with respect to annihilation and pair-creation
processes.
Therefore the chemical potentials of  boson particles $\mu_{p}$ and
boson antiparticles $\mu _{\bar{p}}$ have opposite signs:
\begin{equation}
\mu_{p} \,=\, -\mu_{\bar{p}} \, \equiv \, \mu  \,.
\label{eq:23}
\end{equation}
We are going to consider the system of bosonic particles and bosonic antiparticles
with the conserved density of the isospin number $n_{I} = n^{(-)} - n^{(+)}$,
where $n^{(-)}$ is the particle-number density of bosonic particles and
$n^{(+)}$ is the particle-number density of bosonic antiparticles.
Therefore, the Euler relation includes isospin number density only:
\begin{equation}
\varepsilon \,+\, p \,=\, T\,s \,+\, \mu \, n_{I} \,.
\label{eq:24}
\end{equation}
The total particle-number density is $n = n^{(-)} + n^{(+)}$.
\footnote{The dynamical conservation of the total number of pions in a pion-enriched
system created on an intermediate stage of a heavy-ion collision was considered
in refs.~\cite{kolomeitsev-voskresensky-2018,kolomeitsev-borisov-voskresensky-2018,
kolomeitsev-voskresensky-2019}
}

Roughly speaking, in such a problem the chemical potential  controls the
difference of particle and antiparticle numbers
$\mu \rightarrow (N^{(-)} - N^{(+)})$ whereas the total number of particles is
controlled by the temperature $T \rightarrow (N = N^{(-)} + N^{(+)})$.
Indeed, if some amount of particle-antiparticle pairs $M$ has been created
additionally to the  existing particles $N^{(-)}$ and $N^{(+)}$ in a
closed system, then approximately the same value $\mu $ is in correspondence
$\mu \rightarrow [(N^{(-)} + M) - (N^{(+)} + M)]$ but ${T'}\rightarrow
(N^{(-)} + M + N^{(+)} + M)$, where $T' > T$.
This qualitative consideration indicates the existence of one-to-one correspondence
of independent pairs of variables $(T,\mu) \Leftrightarrow (N,\,N_{I})$.
It is an easy task to show that the latter statement is valid
in ideal quantum gas of particles and antiparticles.
Meanwhile, the rigorous proof of the independence of
thermodynamic variables $n$ and $n_{I}$ in a more general case where the mean
fields, which depend on these variables, are present in the system (see
\cite{anchishkin-gnatovskyy-zhuravel-karpenko-2021}), is not so simple.

In general the mean field $U$ depends on both independent variables $n,\,n_{I}$,
i.e. $U(n,n_{I})$.
On the other hand, as proved in \cite{anchishkin-gnatovskyy-zhuravel-karpenko-2021},
the mean field can be separated into $n$-dependent and $n_{I}$-dependent
pieces where then, it reads respectively for particles and antiparticles as
\begin{eqnarray}
U^{(-)}\big(n,n_{I}\big) &=& U(n) - U_{I}\big(n_{I}\big) \,,
\label{eq:25}
\\
U^{(+)}\big(n,n_{I}\big) &=& U(n) + U_{I}\big(n_{I}\big) \,.
\label{eq:26}
\end{eqnarray}
These signs in eqs.~(\ref{eq:25}) and (\ref{eq:26}) are due to odd dependence on
the isospin number $n_{I}$.

The total pressure in the two-component system reads
\begin{eqnarray}
p &=& -\, g T \int  \frac{d^3k}{(2\pi)^3}
\ln{ \left[ 1 - \exp \left( -\frac{\sqrt{m^2+{\bf k}^2}
+  U(n) - U_{I}(n_{I}) - \mu }{T}\right) \right] } \, -
\nonumber  \\
&& \hspace{1mm}
-\, g T \int  \frac{d^3k}{(2\pi)^3} \ln{ \left[ 1 - \exp \left(
-\frac{\sqrt{m^2+{\bf k}^2}
+ U(n) + U_{I}(n_{I}) + \mu }{T}\right) \right] } \,
+\, P(n,n_{I}) ,
\label{eq:d16}
\end{eqnarray}
where $P(n,n_{I})$ is the excess pressure.\footnote{Here and below we adopt
the system of units $\hbar=c=1$, $k_{_B}=1$}

At the first step of the investigation, we neglect that part of the mean field,
which depends on isospin density, i.e., we assume  $U_{I}(n_{I}) = 0$.
Therefore, in this approximation, the excess pressure also depends only on
the total particle-number density, $P(n)$.

The thermodynamic consistency of the mean-field model can be obtained by putting
in correspondence of two expressions that must coincide in the result.
These expressions, which determine the isospin density, looks like
\begin{equation}
n_{I} \,=\, \left(\frac{\partial p}{\partial \mu}\right)_T \,,
\label{eq:i-density}
\end{equation}
where pressure is given by Eq.(\ref{eq:d16}), and
\begin{equation}
n_{I} \,=\, g \int \frac{d^3k}{(2\pi )^3} \,
\big[ f\big(E(k,n),\mu\big) \,-\, f\big(E(k,n),-\mu\big)\big] \,.
\label{eq:29}
\end{equation}
Here $E(k,n) = \omega_k + U(n)$ with $\omega_k = \sqrt{m^2 + \bs k^2}$ and
the Bose-Einstein distribution function reads
\begin{equation}
f\big(E,\mu\big) = \left[ \exp{ \left( \frac{E - \mu}{T} \right)}  - 1\right]^{-1} \,.
\label{eq:32}
\end{equation}
In order the expressions (\ref{eq:i-density}) and (\ref{eq:29}) to coincide
in the result, the following relation between the mean field and the excess
pressure  arises
\begin{equation}
n\, \frac{\partial U(n)}{\partial n} \,=\, \frac{\partial P(n)}{\partial n} \,.
\label{eq:d20}
\end{equation}
It provides the thermodynamic consistency of the model.
When both components of $\pi^-$-$\pi^+$ system are in the thermal (kinetic)
phase, the  pressure and energy density read
\begin{eqnarray}
\label{eq:30}
p &=& \frac{g}{3} \int \frac{d^3k}{(2\pi )^3}\, \frac{{\bf k}^2}{\omega_k}
\big[ f\big(E(k,n),\mu\big) \,+\, f\big(E(k,n),-\mu\big)\big] \,+\, P(n) \,,
\\
\varepsilon  &=& g \int \frac{d^3k}{(2\pi )^3} \,
E(k,n)\, \big[ f\big(E(k,n),\mu\big) \,+\, f\big(E(k,n),-\mu\big)\big] \,-\, P(n) \,.
\label{eq:31}
\end{eqnarray}
%

\section{Skyrme-like parametrization of the mean field }
\label{sec:skyrme-param}

The thermodynamic mean-field model has been applied for several
physically interesting systems, including the hadron-resonance gas
\cite{anch-vovchenko-2015} and the pionic gas \cite{anch-2016}.
This approach was extended to the case of a bosonic system at $\mu = 0$, which
can undergo Bose condensation
\cite{anchishkin-mishustin-2019, anchishkin-4-2019}.
In the present study, a generalized formalism given in section
\ref{sec:mfm-part-antipart} is used to describe the particle-antiparticle system
of bosons when the  isospin density is kept constant.
As was mentioned  in the previous section,
the mean field in general case splits into two pieces with dependence on the
total particle density $n$ and on the isospin density $n_I$, respectively,
see eqs.~(\ref{eq:25}) and (\ref{eq:26}).
At the first stage of our investigation
we assume that the interaction between particles is described by the Skyrme-like
mean field, which depends only on the total particle-number density $n$.
Loosely speaking, we take into account just a strong interaction.
So, we assume that the mean field reads
\begin{equation}
U(n)\, =\, - \, A\,n\, +\, B\, n^2 \,,
\label{eq:mf1}
\end{equation}
where $A$ and $B$ are the model parameters, which should be specified.
Some additional contribution to the attractive mean field at high
temperatures, ($T \propto 100-160$~MeV), may be provided by other hadrons
present in the system like $\rho$-mesons \cite{shuryak-1991} or
baryon-antibaryon pairs  \cite{Theis}.
As was mentioned in the introduction, an investigation of the properties of
a dense and hot pion gas is well inspired by the formation of the medium with low
baryon numbers at midrapidity what was proved in the experiments
at RHIC, and LHC \cite{adamczyk-2017,abelev-2012}.

For this reason, in our calculations, we consider a general case of $A > 0$,
to study a bosonic system with both attractive and repulsive contributions to
the mean field (\ref{eq:mf1}).
For the repulsive coefficient $B$ we use a fixed value, obtained
from an estimate based on the virial expansion \cite{hansen-2005},
$B = 10 m v_0^2$ with $v_0$ equal to four times the proper volume of
a particle, i.e. $v_0 = 16 \pi r_0^3/3$.
In our numerical calculations we take $v_0 = 0.45$~fm$^3$ that corresponds to
a ``particle radius'' $r_0 \approx 0.3$~fm.
The numerical calculations will be done for bosons with mass $m = 139$~MeV,
which we call conventionally ``pions''.
In this case, the repulsive coefficient is $B/m = 2.025$~fm$^6$, and it is kept
constant through all present calculations.
(For instance, in Ref.~\cite{stashko-anchishkin-2020}authors use the value
$B/m = 21.6$~fm$^6$.)
At the same time, the coefficient $A$, which determines the intensity of
attraction of the mean field (\ref{eq:mf1}), will be varied.
It is advisable to parameterize the coefficient $A$.
We are going to do this with making use of solutions of equation
$U(n) + m = 0$, similar to parametrization adopted in
refs.~\cite{anchishkin-mishustin-2019,anchishkin-4-2019}.
For the given  mean field (\ref{eq:mf1}) there are two roots of this
equation ($n_{1,2} = (A \mp \sqrt{A^2 - 4mB})/2B$)
\begin{equation}
n_1 \, =\, \sqrt{\frac{m}{B}} \left( \kappa - \sqrt{\kappa^2 - 1}\right)\,,
\qquad
n_2 \, =\, \sqrt{\frac{m}{B}} \left( \kappa + \sqrt{\kappa^2 - 1} \right) \,,
\label{eq:n1-n2-10}
\end{equation}
where
\begin{equation}
\kappa \, \equiv\, \frac{A}{2\,\sqrt{m \,B}} \,.
\label{eq:kappa}
\end{equation}
Then, one can parameterize the attraction coefficient as $A = \kappa A_{\rm c}$
with  $A_{\rm c} = 2\sqrt{m B}$.
As we will show below, the dimensionless parameter $\kappa$ is the scale parameter
of the model.
When we fix the isospin density, the parameter $\kappa$ determines the phase
structure of the system.
As it is seen from eq.~(\ref{eq:n1-n2-10}) for the values of parameter $\kappa < 1$
there are no real roots.
The critical value  $A_{\rm c}$ is obtained when both roots coincide, i.e.
when $\kappa = \kappa_{\rm c} = 1$, then $A = A_{\rm c} = 2\sqrt{m B}$.

In general, there are two intervals of the parameter $\kappa$.
1) First interval corresponds to  $\kappa \le 1$,  there are no real roots
of equation $U(n) + m = 0$.
We associate these values of $\kappa$ with a ``weak'' attractive interaction,
and in the present study, we consider variations in the attraction coefficient $A$
for values of $\kappa$ only from this interval.
2) Second interval corresponds to  $\kappa > 1$,  there are two real roots
of equation $U(n) + m = 0$.
We associate this interval with a ``strong'' attractive interaction.
This case will be considered elsewhere.

\medskip

If one assumes a possibility of the Bose-Einstein condensation in the
two-component system, then it is instructive to classify a phase structure of
the system by two basic combinations which determine for the
``weak'' attraction the different thermodynamic states:
(i) Both components, or the boson particles and boson antiparticles, i.e.
$\pi^-$ and $\pi^+$, are in the thermal (kinetic) phase;
(ii) Particles ($\pi^-$) are in the condensate phase, and antiparticles
($\pi^+$) are in the thermal (kinetic) phase - this combination can
be named as the ``cross'' state.

It is necessary to note that the expression ``particles are in the condensate phase''
is, of course, a conventional one.
Because in essence, it is a mixture phase, where at a fixed temperature,
a fraction of $\pi^-$-mesons is in thermal states with momentum $|\bs k| > 0$
and other fraction of this $\pi^-$-component belongs to the Bose-Einstein
condensate, where all $\pi^-$-mesons have zero momentum, $\bs k = 0$.

We are going now to consider  these basic thermodynamic states of the
system using the mean field (\ref{eq:mf1}).

\section{Thermodynamic properties of the boson particle-antiparticle system
under ``weak attraction''}
\label{sec:thrmod-properties}

In the mean-field approach, the behavior of the particle-antiparticle bosonic
system in the thermal (kinetic) phase is determined by the set of two
transcendental equations (we keep $n_I =$~const)
\begin{eqnarray}
\label{eq:tot-n}
n &=&  \int \frac{d^3k}{(2\pi )^3} \,
\left[ f\big(E(k,n),\mu\big) \,+\, f\big(E(k,n),-\mu\big) \right] \,,
\\
n_{I} &=&  \int \frac{d^3k}{(2\pi )^3} \,
\left[ f\big(E(k,n),\mu\big) \,-\, f\big(E(k,n),-\mu\big) \right] \,,
\label{eq:ni}
\end{eqnarray}
where the Bose-Einstein distribution function $f\big(E,\mu\big)$ is defined
in (\ref{eq:32}) and $E(k,n) =  \omega_k + U(n)$.
Equations (\ref{eq:tot-n})-(\ref{eq:ni}) should be solved selfconsistently with
respect to $n$ and $\mu$  for a given temperature $T$ with account for
$n_{I} =$~const.
In the present, we consider the boson system in the Canonical Ensemble,
where the independent canonical variables are $T$ and $n_I$,
particles spin equal to zero.
In this approach, the chemical potential $\mu$ is a thermodynamic quantity that
depends on the canonical variables, i.e., $\mu(T,n_I)$.

In case of the cross state, when the particles, i.e. $\pi^-$-mesons, are in the
condensate phase and antiparticles are still in the thermal (kinetic) phase,
eqs.~(\ref{eq:tot-n}), (\ref{eq:ni}) should be generalized to
include condensate component $n^{(-)}_{\rm cond}$.
Besides this, we should take into account that the particles ($\pi^-$ or
high-density component) can be in the condensed state just under the
necessary condition:
\begin{equation}
U(n) \,-\, \mu  \,=\,  - m \,.
\label{eq:condens-cond}
\end{equation}
As the temperature decreases from high values, when both $\pi^-$ and
$\pi^+$ are in the thermal phase, the density of the
$\pi^-$-component, namely $n^{(-)}(T,\mu)$, crosses the critical curve at
the temperature $T_{\rm c}^{(-)}$, where the condition (\ref{eq:condens-cond})
is satisfied.
The latter  means that the curve $n^{(\rm id)}_{\rm lim}(T)$, which is
defined as
\begin{equation}
n^{(\rm id)}_{\rm lim}(T)  \,=\,  \int \frac{d^3k}{(2\pi)^3}\,
f\big(\omega_k,\mu\big)\Big|_{\mu = m} \,,
\label{eq:nlim-id}
\end{equation}
is  the critical curve for $\pi^-$-mesons or the high-density component.
Here $f(\omega_k,\mu)$ is the Bose-Einstein distribution function defined in
(\ref{eq:32}).
As we see function (\ref{eq:nlim-id}) represents the maximal density of thermal
(kinetic) boson particles of the ideal gas at temperature $T$ when $\mu = m$.
Hence, we obtain that the critical curve in the mean-field approach under
consideration for the boson particles coincides with the critical curve for
the ideal gas.

With account for eqs.~(\ref{eq:condens-cond}) and (\ref{eq:nlim-id}) we write
the generalization of the set of eqs.~(\ref{eq:tot-n}), (\ref{eq:ni})
\begin{eqnarray}
\label{eq:tot-n2a}
n &=&  n^{(-)}_{\rm cond}(T) + n^{(\rm id)}_{\rm lim}(T) \,
+\, \int \frac{d^3k}{(2\pi )^3} \,f\big( E(k,n),-\mu \big) \,,
\\
n_{I} &=&  n^{(-)}_{\rm cond}(T) + n^{(\rm id)}_{\rm lim}(T)
- \int \frac{d^3k}{(2\pi )^3} \, f\big(E(k,n),-\mu\big)  \,.
\label{eq:ni2a}
\end{eqnarray}
Meanwhile, using relation (\ref{eq:condens-cond}) between the mean field and
the chemical potential, this set of equations can be reduced to just one
equation for $n^{(+)}$:
%
\begin{equation}
n^{(+)} \,=\, \int \frac{d^3k}{(2\pi )^3} \,
f\big( E(k,n),-\mu \big)\Big|_{\mu = U(n) + m}  \,,
\label{eq:nneg-ST2}
\end{equation}
where $U(n) = U\!\left( 2n^{(+)} + n_I \right) $ and
$E(k,n) = \omega_k + U\!\left( 2n^{(+)} + n_I \right)$.
Solution of  eq.~(\ref{eq:nneg-ST2}) for temperatures $T$ from the interval
$T < T_{\rm c}^{(-)}$ provides the density $n^{(+)}(T)$ of $\pi^+$ mesons.

One can see from eqs.~(\ref{eq:tot-n2a}), (\ref{eq:ni2a}) that the
particle-number density $n^{(+)}$ is provided only by thermal $\pi^{+}$ mesons.
Whereas, the density $n^{(-)}$ is provided by two fractions:
the  condensed particles ($\pi^{-}$ mesons at $\bs k = 0$) with the
particle-number density $n^{(-)}_{\rm cond}(T)$, and
thermal $\pi^{-}$ mesons  at $|\bs k| > 0$ with the particle-number density
$n^{(\rm id)}_{\rm lim}(T)$.
The particle-density sum rule for these phase of $\pi^{-}$ mesons
in the interval $T < T_{\rm c}^{(-)}$ reads
\begin{equation}
n^{(-)} \,=\, n^{(-)}_{\rm cond}(T) \,+\, n^{(\rm id)}_{\rm lim}(T)  \,.
\label{eq:nneg-tot}
\end{equation}
%

\subsection{Numerical results}
\label{sec:numeric-calc}

At high temperatures, i.e. $T \ge T_{\rm c}^{(-)}$,
both components of the bosonic particle-antiparticle system are in the thermal
phase and thermodynamic properties of the system are determined by
the set of eqs.~(\ref{eq:tot-n}) and (\ref{eq:ni}).
Solving this set for  given values $T$ and $n_I$ we obtain the functions
$\mu(T,n_I)$ and $n(T,n_I)$ and then other thermodynamic quantities.

When we decrease temperature, after crossing the value $T = T_{\rm c}^{(-)}$ the
particles which belong to the high-density component (or $\pi^{-}$-mesons)
start to ``drop down'' into the condensate state, which is characterized
by the value of momentum $\bs k = 0$.
In the limit, when $T = 0$, all particles of the high-density component,
i.e. $\pi^{-}$-mesons, are in condensed state and $n^{(-)} = n_I$.
At the same time, the particles of the low-density component or $\pi^{+}$-mesons
being in the thermal phase lose the density $n^{(+)}$ with a decrease of
temperature, and it becomes rigorously zero at $T = 0$.
For the  temperature interval  $T < T_{\rm c}^{(-)}$ equations
(\ref{eq:tot-n}), (\ref{eq:ni}) should be generalized
and now thermodynamic properties of the system are determined by
eq.~(\ref{eq:nneg-ST2}), where we take into account that $\mu = - U(n) + m$
for all temperatures of this interval unless the high-density
component $n^{(-)}$ is in condensed state.
Otherwise it is necessary to solve the set of eqs.~(\ref{eq:tot-n}) and
(\ref{eq:ni}) for the region where $n^{(-)}$ appears again in the thermal
(kinetic) phase.

\begin{figure}
\centering
\includegraphics[width=0.49\textwidth]{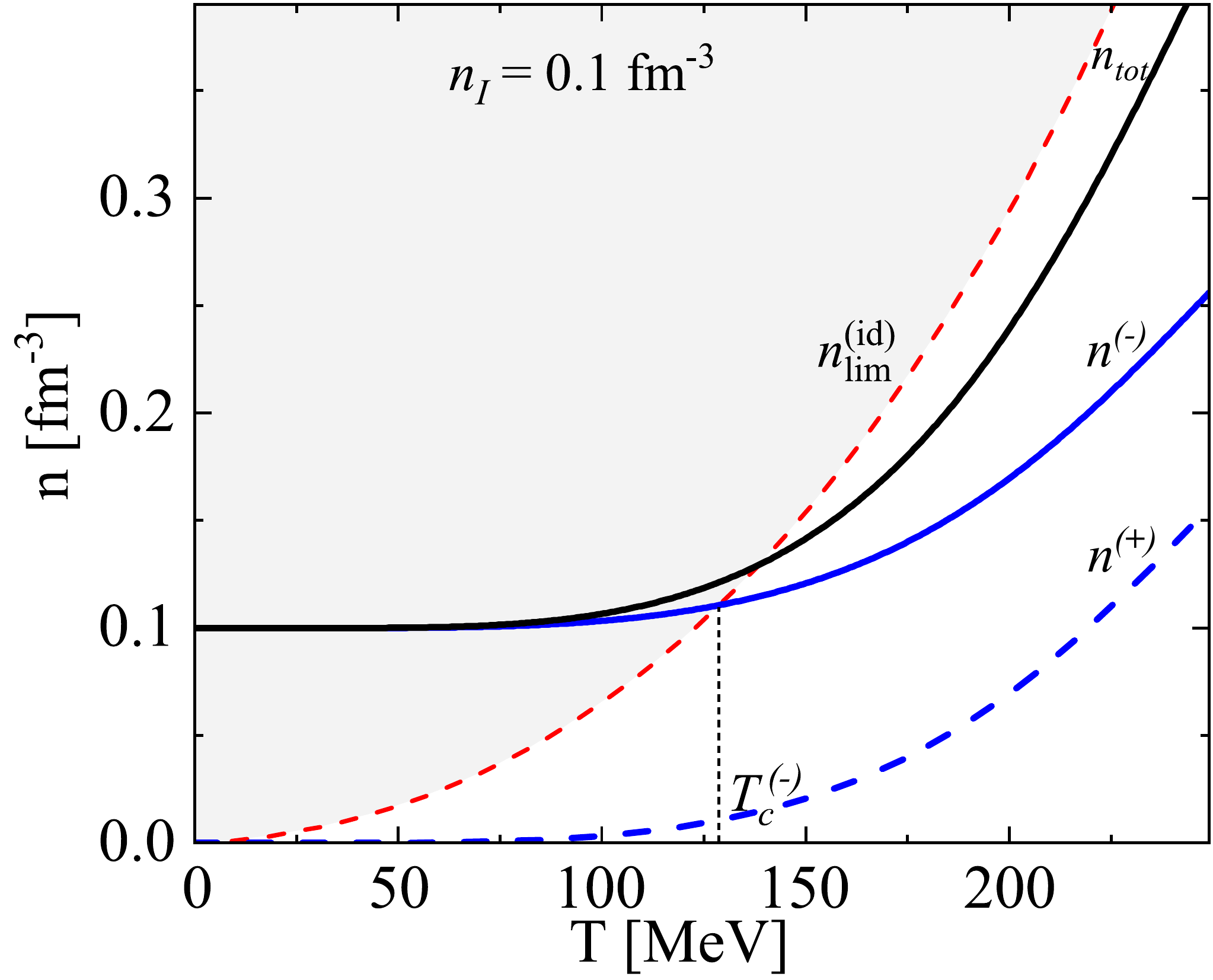}
\includegraphics[width=0.49\textwidth]{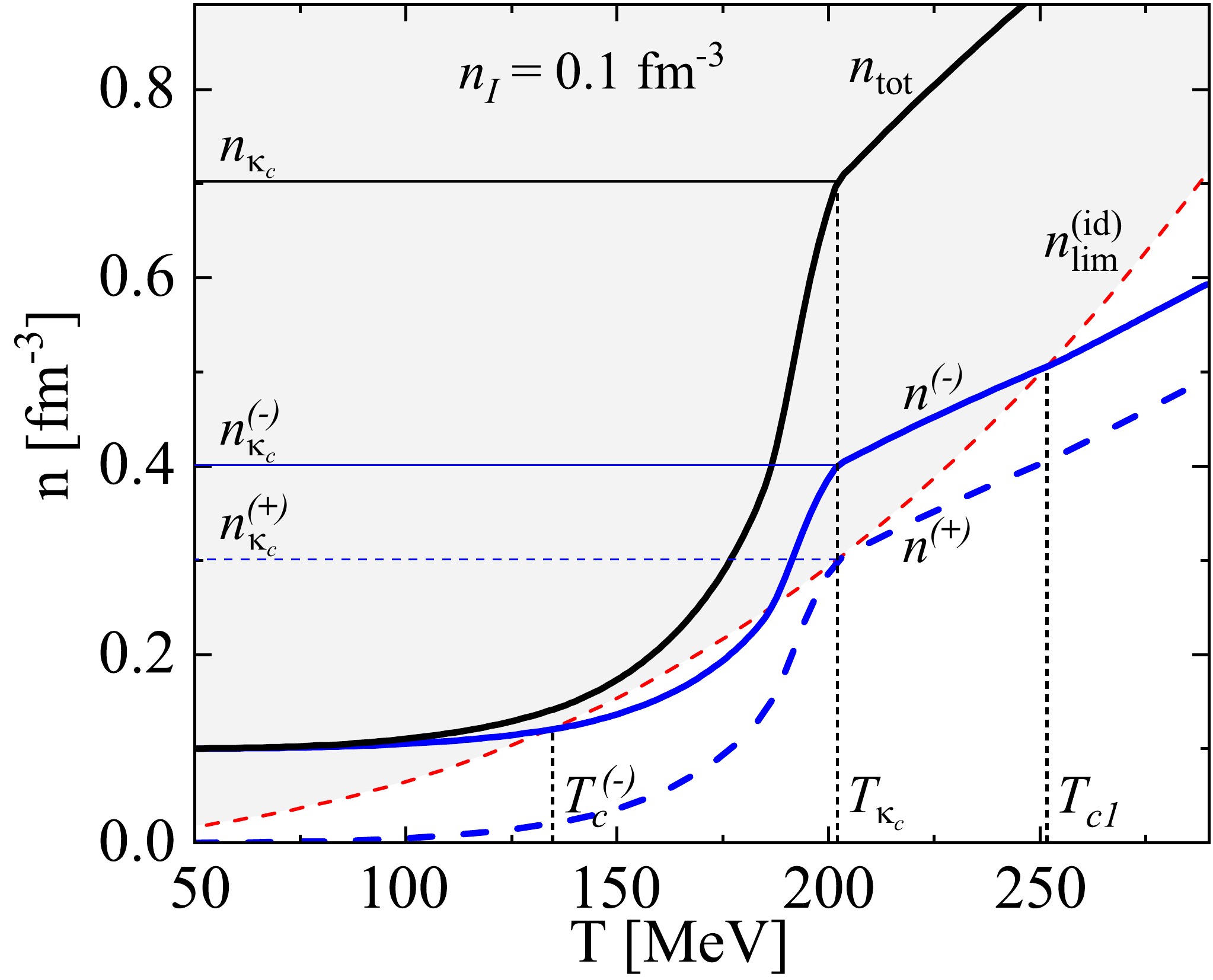}
\caption{
{\it Left panel:} The particle-number densities $n^{(+)}$, $n^{(-)}$ and
$n_{\rm tot} = n^{(+)} + n^{(-)}$ versus temperature for the
interacting $\pi^+$-$\pi^-$ pion gas in the mean-field model.
The total isospin density is kept constant, $n_{\rm I} =0.1$~fm$^{-3}$, and
the attraction parameter is $\kappa = 0.5$.
The maximum density $n_{\rm lim}^{\rm (id)}$ of the ideal gas
of thermal pions at $\mu = m_\pi$ is shown by the red dashed line.
The shaded area shows the possible states of condensed particles.
The Bose-Einstein condensation of $\pi^-$ mesons occurs at the temperature
$T_{\rm c} = T_{\rm c}^{(-)}$.
{\it Right panel:} The same as on the left panel, but with the parameter
$\kappa = \kappa_{\rm c} = 1 \,.$
Here $n_1 = n_2 \equiv n_{\kappa_{\rm c}}$ (see eq.~(\ref{eq:n1-n2-10})),
$n_{\kappa_{\rm c}}^{(-)} = \left(n_{\rm tot} + n_I\right)/2$,
$n_{\kappa_{\rm c}}^{(+)} = \left(n_{\rm tot} - n_I\right)/2$ and
$T_{\kappa_{\rm c}}$ is the temperature at which the curve
$n^{(+)}(T)$ touches the critical curve $n_{\rm lim}^{\rm (id)}$.
 }
\label{fig:npos-nneg-nq01-k05-1}
\end{figure}

For parameters $n_{\rm I} =0.1$~fm$^{-3}$, $\kappa = 0.5$ and $\kappa = 1.0$
we solve the set of eqs.~(\ref{eq:tot-n}), (\ref{eq:ni}) for the thermal
phase and eq.~(\ref{eq:nneg-ST2}) for the ``cross'' thermodynamic state.
The behavior of the density $n^{(+)}$ of $\pi^{+}$-mesons and  the density
$n^{(-)}$ of $\pi^{-}$-mesons are depicted
in Fig.~\ref{fig:npos-nneg-nq01-k05-1}.
In this figure, we also depicted the behavior of the total density of mesons
$n = n^{(+)} + n^{(-)}$ depending on temperature (in the field of the figure,
this density is denoted as $n_{\rm tot}$).

Analyzing the behavior of the condensate creation
(see Fig.~\ref{fig:npos-nneg-nq01-k05-1}), it is necessary to note that just
the high-density component of the particle-antiparticle gas undergoes the phase
transition to the Bose-Einstein condensate.
If we apply our consideration to pion gas with $n_I = n_\pi^{(-)} - n_\pi^{(+)} > 0$
this means that $\pi^{-}$-component undergoes the phase transition to
the Bose-Einstein condensate and the low-density component or $\pi^{+}$ mesons
exist only in the thermal phase for the whole range of temperatures.
%
%
Hence, it makes sense to look at $T_{\rm c} = T_{\rm c}^{(-)}$ for the
Bose-Einstein condensate of $\pi^{-}$ mesons only in the lattice calculations and
in an experiment, for instance, in heavy-ion collisions.

At the same time, the temperature behavior of the particle-number density $n^{(+)}$
(see Fig.~\ref{fig:npos-nneg-nq01-k05-1})
is very similar to the behavior of the pion density $n(T)$ for $\kappa \le 1$
obtained in Ref.~\cite{anchishkin-mishustin-2019}, where the pion system at
$\mu = 0$ was investigated.
Note that we consider the system of pions only for ``weak'' attraction in
the present study, i.e., at $\kappa \le 1$.
As was shown in Ref.~\cite{anchishkin-mishustin-2019} the behavior of the pion
system at $\kappa > 1$ is drastically different.
In this case, with an increase in temperature at $T = T_{\rm cd} < T_{\rm c}$,
the system undergoes the first-order phase transition.

\subsubsection{The critical temperature}
\label{sec:crit-temp}

Equation (\ref{eq:tot-n2a}) can be used to determine the critical
temperature $T_{\rm c}^{(-)}$.
Indeed, let us take into account that at the crossing point with the critical
curve the density of condensate  is zero so far,
$n^{(-)}_{\rm cond}\big(T_{\rm c}^{(-)}\big) = 0$,
and the density of thermal $\pi^-$ particles becomes equal to
$n^{(-)}\big(T_{\rm c}^{(-)}\big) = n^{(\rm id)}_{\rm lim}\big(T_{\rm c}^{(-)}\big)$.
Then, at this temperature $T = T_{\rm c}^{(-)}$ on the l.h.s. of
eq.~(\ref{eq:tot-n2a}) we have
$n = 2 n^{(\rm id)}_{\rm lim}\big(T_{\rm c}^{(-)}\big) - n_I$,
and now at this temperature point on the critical curve eq.~(\ref{eq:tot-n2a})
with respect to $T$ reads as:
\begin{equation}
n^{(\rm id)}_{\rm lim}(T) - n_I = \int \frac{d^3k}{(2\pi )^3} \,f\big( E(k,n),-\mu
\big)\Big|_{\mu = U(n) + m}
\quad {\rm with}  \quad
E(k,n) = \omega_k + U\!\left( 2n^{(\rm id)}_{\rm lim} - n_I \right) \,.
\label{eq:nneg-ST2-2}
\end{equation}
Solving eq.~(\ref{eq:nneg-ST2-2}) at $n_I = 0.1$~fm$^{-3}$, for $\kappa = 0.5$ and
$\kappa = \kappa_{\rm c} = 1$ we obtained $T_{\rm c}^{(-)} = 129$~MeV and
$T_{\rm c 1}^{(-)} = 251$~MeV, respectively.
These results are depicted in Fig.~\ref{fig:npos-nneg-nq01-k05-1} in the left
and right panels, respectively.

\begin{figure}
\centering
\includegraphics[width=0.49\textwidth]{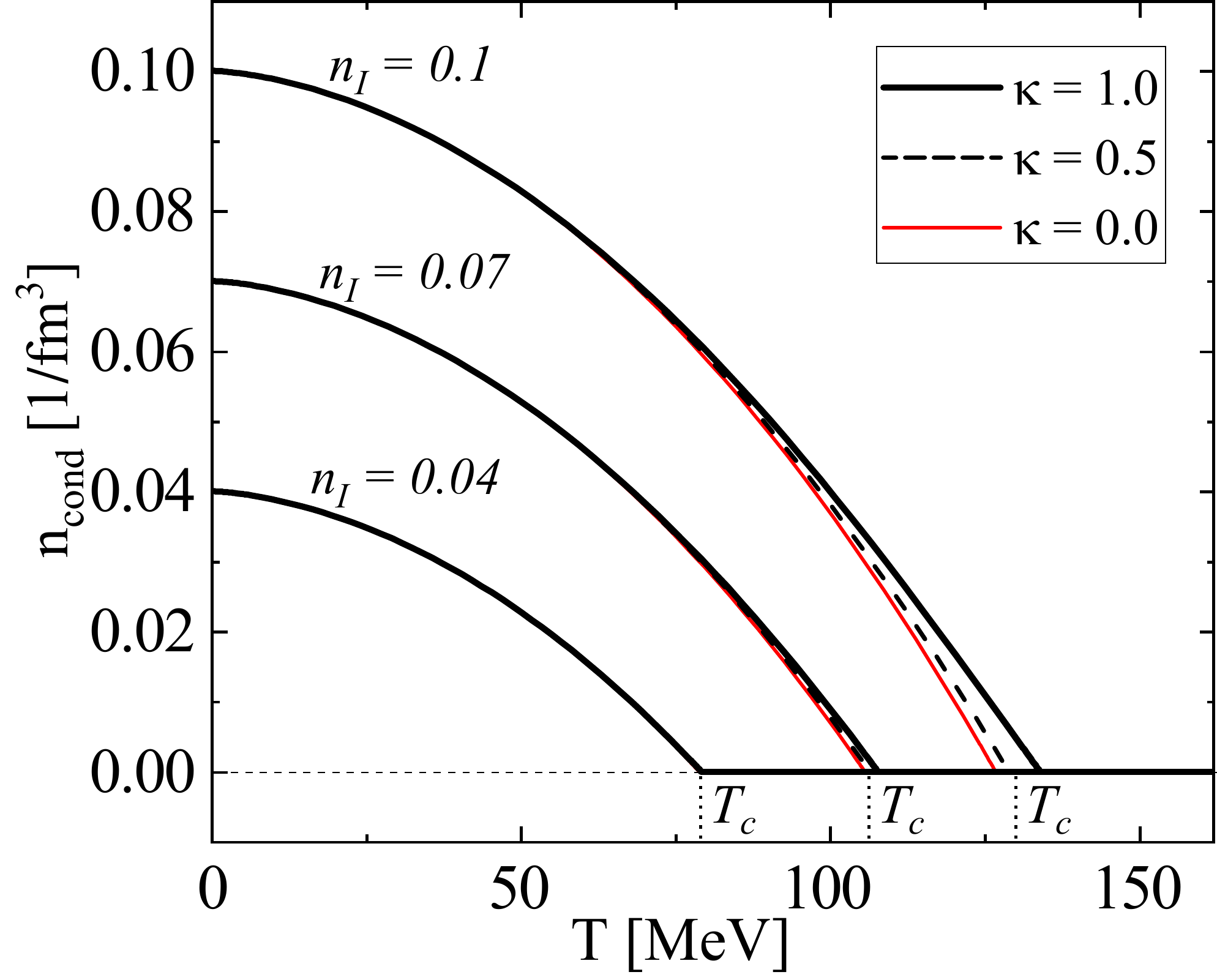}
\includegraphics[width=0.49\textwidth]{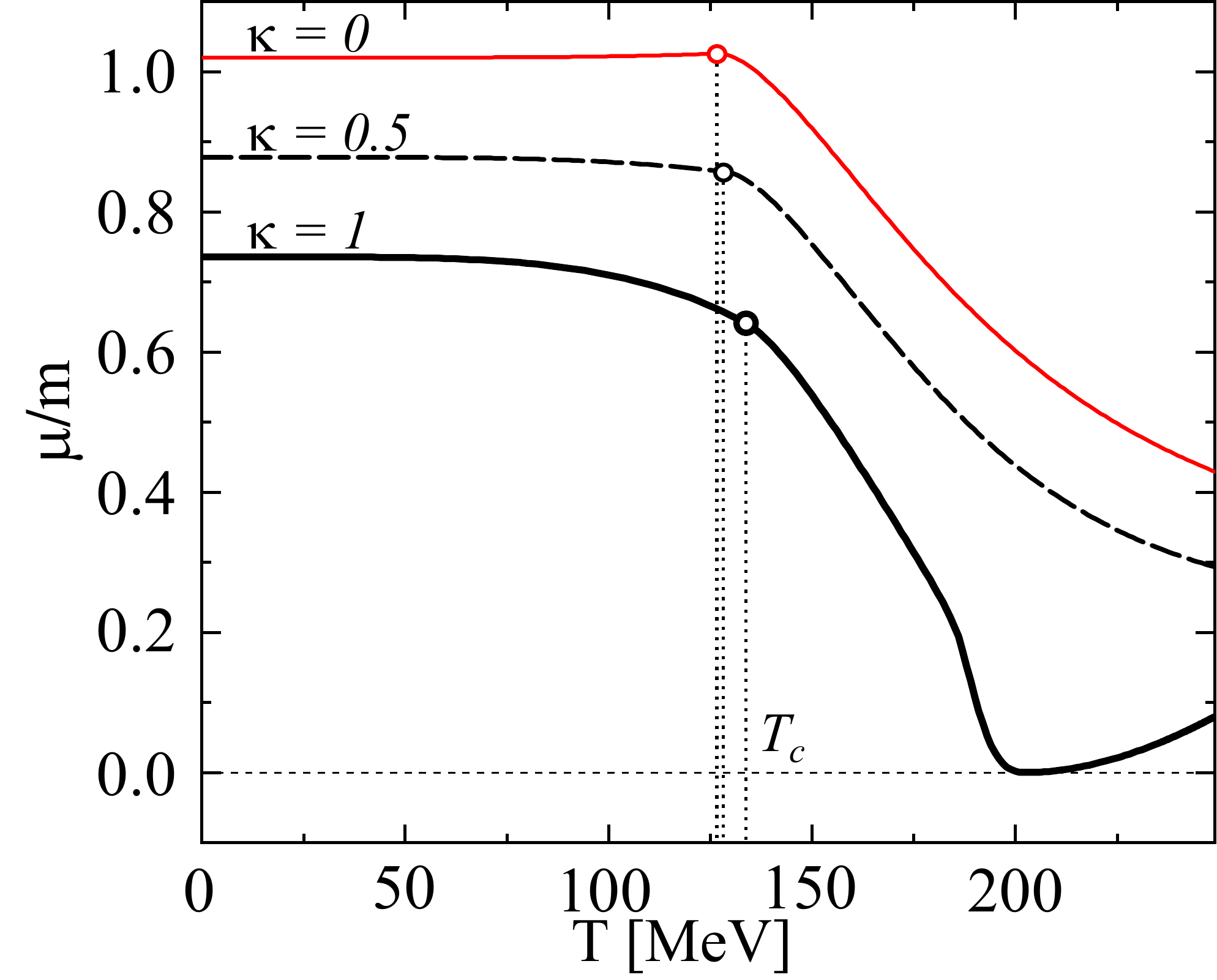}
\caption{ {\it Left panel:} The density of condensate versus temperature in the
particle-antiparticle self-interacting system for three values of the isospin
density, $n_{\rm I} = 0.04,\, 0.07,\, 0.1$~fm$^{-3}$.
\\
{\it Right panel:} The chemical potential versus temperature at values of
the attraction parameter $\kappa = 0.0,\, 0.1,\, 0.5,\, 1.0$ and
the isospin density $n_{\rm I} = 0.1$~fm$^{-3}$.
The marked points on the curves correspond to the critical temperature
$T_{\rm c}^{(-)}$.
In both panels we set $T_{\rm c} = \left\langle T_{\rm c}^{(-)} \right\rangle$.
}
\label{fig:2nd-order}
\end{figure}

It turns out that $T_{\rm c}^{(-)}$ is the critical temperature, which
determines the phase transition with the formation of a BEC for the entire
pion system since antiparticles or $\pi^+$-mesons, which represent the
low-density component $n^{(+)}(T)$, are entirely in a thermal state for all
temperatures.
Thus, condensate is created only by particles or by $\pi^-$-mesons,
i.e., $n_{\rm cond} = n_{\rm cond}^{(-)}$, and this particle-number density
plays the role of the order parameter.

The condensate densities as functions of temperature obtained in the framework
of our model for three values of the attraction parameter,
$\kappa = 0.0,\, 0.5,\, 1.0$, and for three values of the isospin density,
$n_{\rm I} =0.04,\, 0.07,\, 0.1$~fm$^{-3}$, are depicted
in Fig.~\ref{fig:2nd-order}, left panel.
We record a minimal difference in the critical temperature $T_{\rm c}^{(-)}$
when the attraction parameter $\kappa$ changes, the difference does not exceed
$4$~MeV when $n_{\rm I} = 0.1$~fm$^{-3}$.
This difference is much less, as we can see in Fig.~\ref{fig:2nd-order} for
smaller isospin densities.
Then it would be helpful to define only one average value of $T_{\rm c}^{(-)}$ as
\begin{equation}
T_{\rm c} \,=\, \left\langle T_{\rm c}^{(-)} \right\rangle\,.
\label{eq:tc}
\end{equation}
For example for $n_{\rm I} = 0.1$~fm$^{-3}$ the averaging gives
$T_{\rm c} \approx 129$~MeV.
The temperature $T_{\rm c}$  ``signals'' the creation of condensate
when temperature decreases and crosses this value.
Note that the critical temperature $T_{\rm c}$ is practically independent of
the attraction parameter $A$ of the mean field (\ref{eq:mf1}).
In other words, the average attraction between particles in the system has
little effect on the critical temperature.

The dependence of the chemical potential on temperature is depicted
in Fig.~\ref{fig:2nd-order} in the right panel for three values of the
attraction parameter, $\kappa = 0.0,\, 0.1,\, 0.5,\, 1.0$.
First of all, we notice that the chemical potential is almost independent of
temperature when condensate exists in the system, i.e., in the interval
$0 < T \le T_{\rm c}$.
Value of $\mu$ changes from $1.02 m_\pi$ at the absence of attraction, $\kappa = 0.0$,
to $\mu = 0.74 m_\pi$ for the critical attraction parameter $\kappa = 1.0$.
Hence, for $0 \le \kappa \le 1$ the chemical potential is in the range
$103 \le \mu \le 142$~MeV.
It is intriguing to remind that already first attempts to fit the $p_{\rm T}$
spectra of $\pi^{-}$-mesons in O+Au collisions at $200$~AGeV/nucleon
(at midrapidity) by the ideal-gas Bose-Einstein distribution results in the values
$\mu \approx 126$~MeV,  $T \approx 167$~MeV
and in S+S collisions at $200$~AGeV/nucleon it results in the values
$\mu \approx 118$~MeV,  $T \approx 164$~MeV \cite{kataja-ruuskanen-1990}.
So, the fit of data
required the pion chemical potential in the range $\mu \approx 115-130$~MeV
what we can  formally compare with the values of the chemical potential
obtained in our model.

\subsubsection{The heat capacity}
\label{sec:crit-temp}

The derivative of the chemical potential on temperature has a jump in points
marked on the curves as small black circles, see Fig.~\ref{fig:2nd-order}
right panel.
These points on the curves $\mu(T)$ correspond to $T_{\rm c}^{(-)}$,
which values differ from one another not more than $\Delta T = 4$~MeV.
As we concluded before, this is the temperature of phase transition,
see eq.~(\ref{eq:tc}), which practically does not depend on the intensity of
attraction.
To prove that this is indeed a phase transition of the second order, we first
calculate the heat capacity $c_{\rm v}$ as
\footnote{As a matter of fact, here we calculate the volumetric heat capacity,
which is  the heat capacity $C_{\rm V}$ of a system  divided by the volume $V$, i.e
$c_{\rm v} = C_{\rm V}/V$.
}
\begin{equation}
c_{\rm v} \,=\, - T \,  \frac{\partial^2 f}{\partial T^2} \,,
\label{eq:specific-heat}
\end{equation}
where $f(T,n_I) = - p(T,n_I) + n_{I} \mu(T,n_I)$ is the density of free energy.
We are going to calculate $f(T,n_I)$ for two thermodynamic scenarios,
when $T \ge T_{\rm c}$ and when $T < T_{\rm c}$.

Having solved  eqs.~(\ref{eq:tot-n}), (\ref{eq:ni}) then, using
eq.~(\ref{eq:30}) one can calculate pressure for the case when particles
and antiparticles are both in the thermal phase, i.e. $T \ge T_{\rm c}$.
In this case, the density of free energy looks like
\begin{equation}
f \,=\,  n_{I}\, \mu(T,n_I) \,
-\, \frac 13 \int \frac{d^3k}{(2\pi )^3} \frac{{\bf k}^2}{\omega_k}
\left[ f\big(E(k,n),\mu\big) \,+\, f\big(E(k,n),-\mu\big) \right]  \,-\, P(n) \,,
\label{eq:free-energy-density-r}
\end{equation}
where functions $n(T,n_I)$ and $\mu(T,n_I)$ are known.
The excess pressure $P(n)$ is obtained by integrating eq.~(\ref{eq:d20}) for
the Skyrme-like parametrization of the mean field (\ref{eq:mf1}):
\begin{equation}
P(n) \,=\, - \frac A2 \, n^2 \,+\, \frac{2B}{3} \, n^3 \,,
\label{eq:pex-skyrme}
\end{equation}
where $P(n = 0) = 0$ is taking into account.

\begin{figure}
\centering
\includegraphics[width=0.49\textwidth]{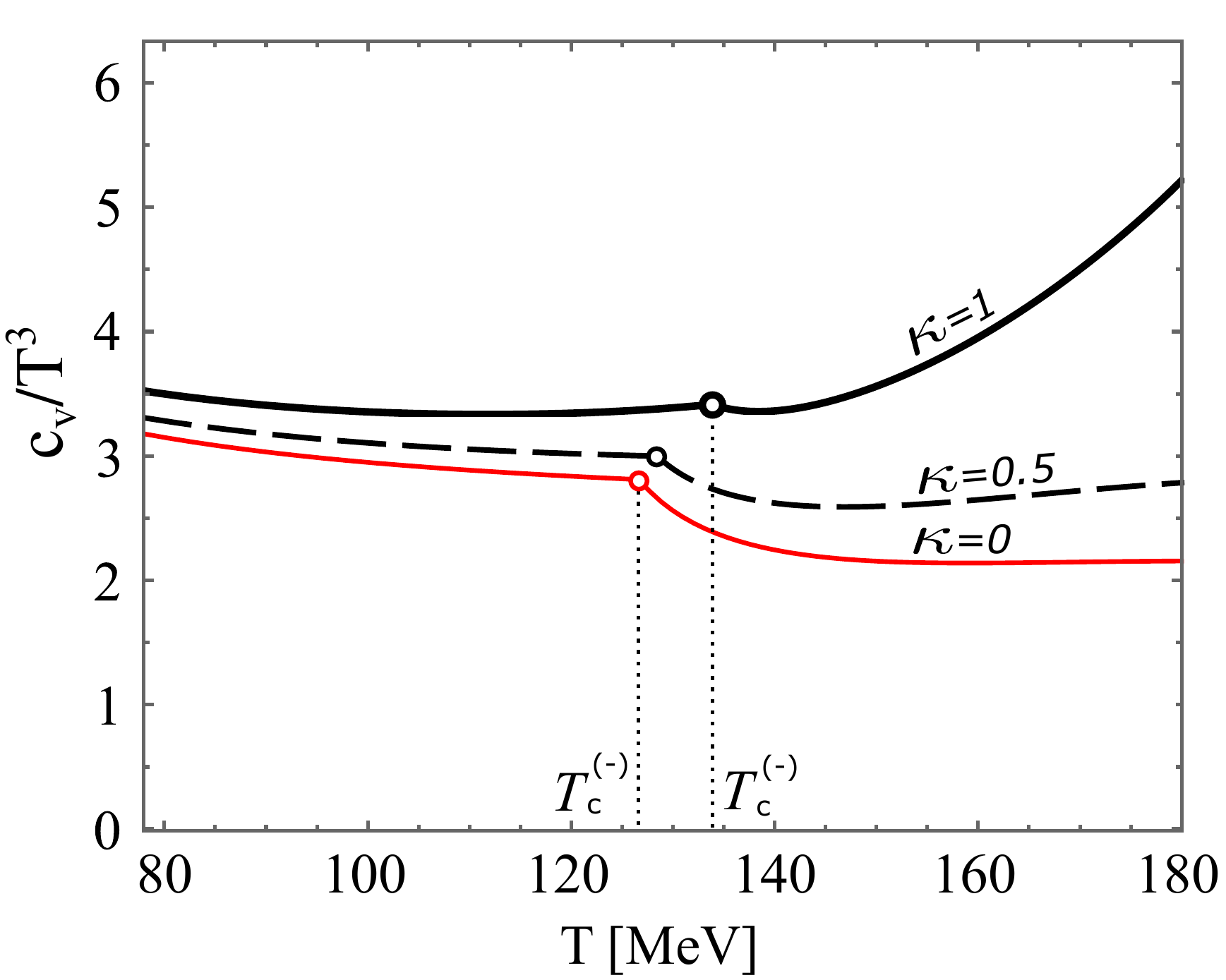}
\includegraphics[width=0.49\textwidth]{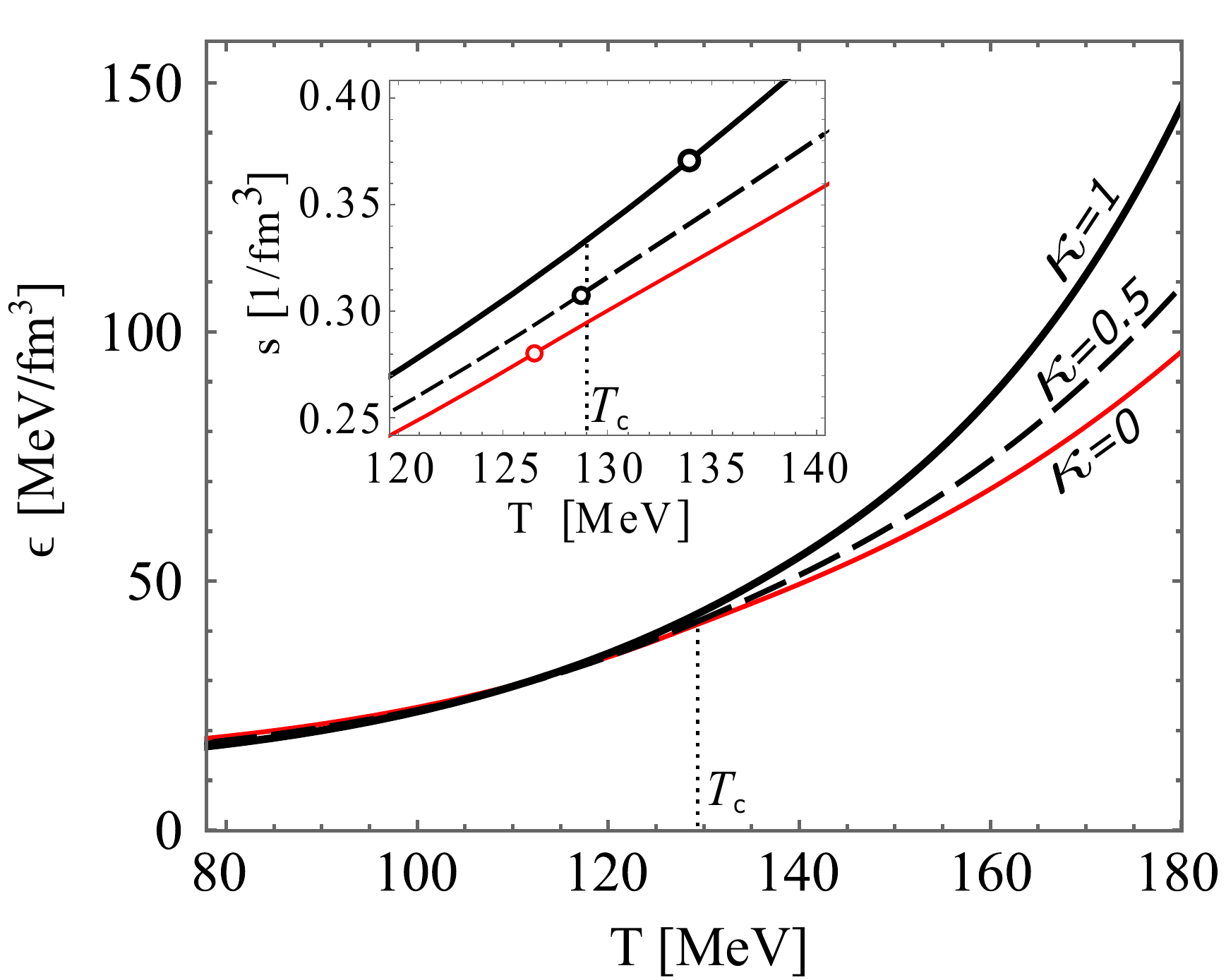}
\caption{ {\it Left panel:}
Heat capacity normalized to $T^3$ as a function of temperature in a
self-interacting $\pi^- - \pi^+$ meson system.
The isospin density is kept constant, $n_{\rm I} = 0.1$~fm$^{-3}$.
The curves are marked with the attraction parameter $\kappa$.
{\it Right panel:} Energy density versus temperature for
the same meson system and the same conditions as in the left panel.
The entropy density versus temperature in the vicinity of $T_{\rm c}$
is shown in a small window for attraction parameters
$\kappa = 0,\, 0.5,\, 1.0$.
We set $T_{\rm c} = \left\langle T_{\rm c}^{(-)} \right\rangle$.
}
\label{fig:cv-s}
\end{figure}

For temperatures less than $T_{\rm c}$, when the high-density component of
the pion gas ($\pi^-$ mesons) is in the condensate phase, and the low-density
component ($\pi^+$ mesons) is in the thermal phase, the density of  free
energy reads
\begin{equation}
f \,=\,  n_{I}\, [U(n) + m]
- \frac 13 \int \frac{d^3k}{(2\pi )^3} \frac{{\bf k}^2}{\omega_k}
f\big(\omega_k,\mu\big)\Big|_{\mu = m}
- \frac 13 \int \frac{d^3k}{(2\pi )^3} \frac{{\bf k}^2}{\omega_k}
f\big(E(k,n),-\mu\big)\Big|_{\mu = U(n) + m} \,- P(n) .
\label{eq:free-energy-density-l}
\end{equation}
Here the total pion density is $n = 2n^{(+)} + n_I$, $\mu = U(n) + m$ as in
eq.~(\ref{eq:nneg-ST2}), $E(k,n) = \omega_k + U(n)$ and $n^{(+)}(T,n_I)$ is
solution of eq.~(\ref{eq:nneg-ST2}).

Using the density of  free energy (\ref{eq:free-energy-density-r}) to the
right of $T_{\rm c}$ and  (\ref{eq:free-energy-density-l}) to the left of
$T_{\rm c}$, respectively, we calculate the  heat capacity normalized to $T^3$,
as function of temperature at $n_{\rm I} =0.1$~fm$^{-3}$ for three values of the
attraction parameter $\kappa = 0,\, 0.5,\, 1.0$.
These dependencies are depicted in Fig.~\ref{fig:cv-s} in the left panel.
The temperature dependence of the heat capacity is a continuous function.
However,
the derivative of this function has a finite discontinuity, which indicates a
second-order phase transition, where the condensate density is an order
parameter (strictly speaking, this is a third-order phase transition).
To make sure that this is indeed a second-order phase transition without the
release of latent heat at the temperature $T_{\rm c}$, we calculate the energy
density $\varepsilon$ for the same set of parameters $\kappa$, the functions
$\varepsilon(T)$ are shown in Fig.~\ref{fig:cv-s} in the right panel.
To be sure that the first derivative of the free energy is a smooth function,
we calculate the entropy density $s = - \partial f(T,n_I)/\partial T$, its
dependence on temperature in the vicinity of $T_{\rm c}$ is shown in a small
window in Fig.~\ref{fig:cv-s} on the right panel for three values of the
attraction parameter $\kappa = 0,\, 0.5,\, 1.0$.
Indeed, one can see that the temperature dependencies of the energy density
and entropy density are continuous and smooth at $T = T_{\rm c}$, which proves
that the system undergoes a second-order phase transition at this temperature.
It is also interesting to note that the energy density in the temperature
interval $0 < T \le T_{\rm c}$ is practically independent of the ``weak''
attraction ($0 \le \kappa \le 1$) between the particles.

\medskip

\begin{figure}
\centering
\includegraphics[width=0.32\textwidth]{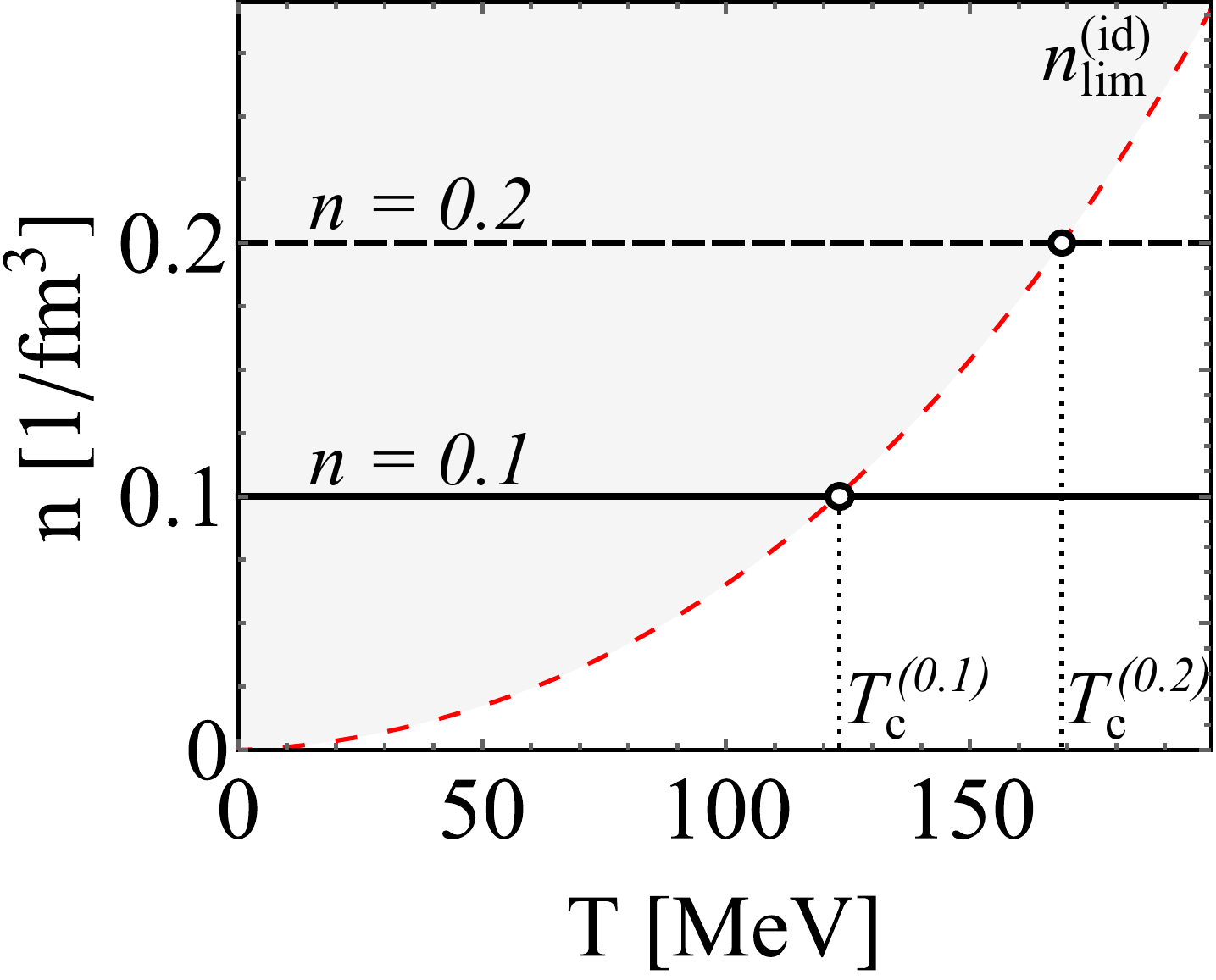}
\includegraphics[width=0.301\textwidth]{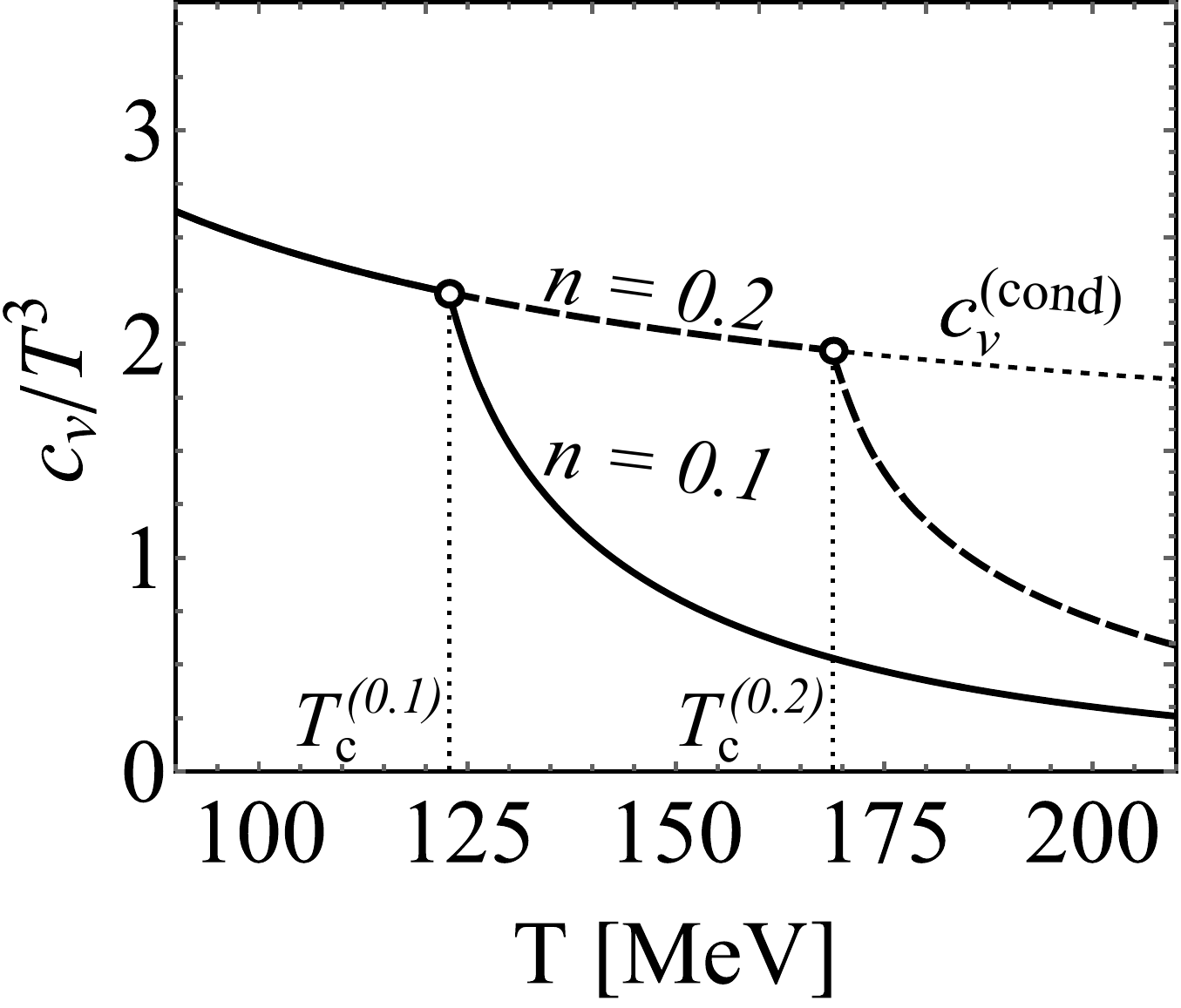}
\includegraphics[width=0.343\textwidth]{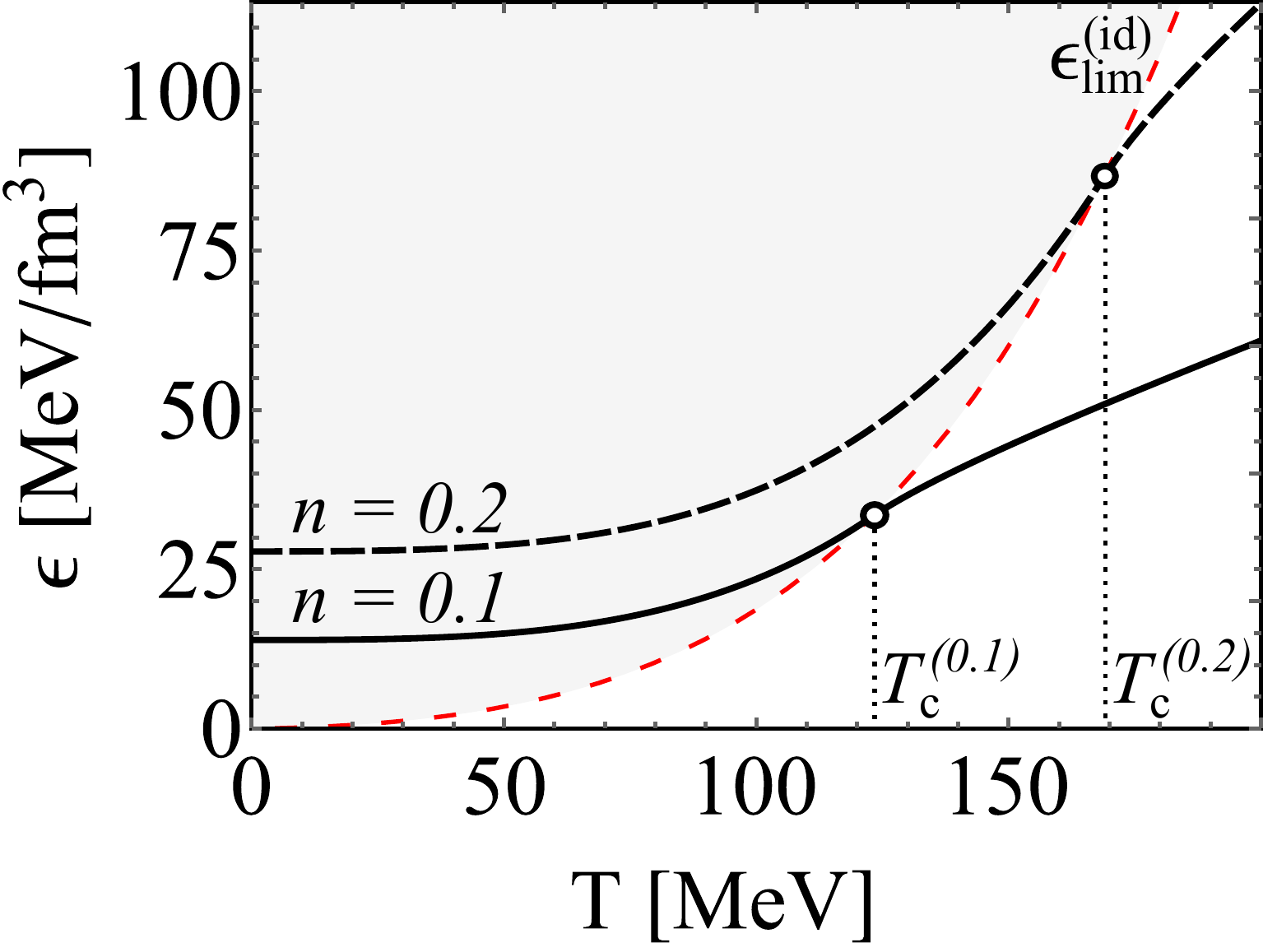}
\caption{
{\it Left panel:}
Particle-number density versus temperature in the ideal single-component gas.
The horizontal lines represent two constant particle density samples,
$n = 0.1,\, 0.2$~fm$^{-3}$, which correspond to critical temperatures
$T_{\rm c}^{(0.1)}$, $T_{\rm c}^{(0.2)}$, respectively.
{\it Middle panel:}
Heat capacity normalized to $T^3$ as a function of temperature in the
ideal single-component gas where the particle-number density is kept constant.
%
{\it Right panel:} Energy density versus temperature for the same system and
conditions as in the left panel.
The red dashed line marked as $\epsilon_{\lim}^{\rm (id)}$ represents the energy
density of the states that belong to the critical curve $n_{\lim}^{\rm (id)}$
depicted in the left panel.
}
\label{fig:cv-ideal}
\end{figure}

We will now fix some similarities between the picture obtained above for
the interacting two-component particle-antiparticle system when $n_I =$~const
and the single-component ideal gas, where we keep constant the particle-number
density $n$.
First, the behavior of the high-density component, in the ``condensate''
temperature interval $T \le T_{\rm c}$ in the system with interaction,
is similar to the behavior of the single-component ideal gas ($m = m_\pi$)
when $n_I = n =$~const.
Indeed, that is seen when one compares the dependance $n = n^{(-)}(T)$
in Fig.~\ref{fig:npos-nneg-nq01-k05-1} and dependence $n =$~const
in Fig.~\ref{fig:cv-ideal} on left panel.
Next, we compare the heat capacities in these two boson systems,
a question of particular interest is the behavior of the heat capacity
at the critical temperature.
%
%
For the ideal gas, it is natural to treat the problem in
the Canonical Ensemble, where the canonical variables are $T$ and $n$.
The critical temperature $T_{\rm c}$ is the starting point for the
onset of condensation when the temperature is decreasing.
For a given density $n$ the critical temperature can be determined as solution
of the transcendental equation $n = n_{\rm lim}^{\rm (id)}(T_{\rm c})$, where
$n_{\rm lim}^{\rm (id)}(T)$ is defined in (\ref{eq:nlim-id}).

The energy density in the condensate phase consists of two contributions,
for the ideal gas it reads
\begin{equation}
\varepsilon \,=\,  m \, n_{\rm cond}(T) \,+\, \int \frac{d^3k}{(2\pi )^3} \, \omega_k \,
f\big(\omega_k,\mu\big)\Big|_{\mu = m} \,,
\label{eq:ideal-energy-density}
\end{equation}
where $n_{\rm cond}(T) = n - n_{\rm lim}^{\rm (id)}(T)$.
We calculate the heat capacity $c_{\rm v} = \partial \varepsilon(T,n)/\partial T$,
which is attributed to the condensate phase, and obtain
\begin{equation}
c_{\rm v}^{\rm (cond)} \,=\,  \frac{1}{4T^2} \,\int \frac{d^3k}{(2\pi )^3} \,  \,
\left[\frac{E_{\rm kin}}{\sinh{(E_{\rm kin}/T)}} \right]^2\,,
\label{eq:cv-ideal-cond}
\end{equation}
where $E_{\rm kin} = \omega_k - m$ is the single-particle kinetic energy.
One can see that the dependence of the heat capacity $c_{\rm v}^{\rm (cond)}(T)$
of the ideal gas in the condensate phase has a universal
character, it limited on the right end by the value of $T_{\rm c}$, which
in turn, depends on the given particle-number density $n$.
This feature is seen in Fig.~\ref{fig:cv-ideal} in the middle panel, where we
consider two samples of the density $n = 0.1,\, 0.2$~fm$^{-3}$.
It is evidently seen that the derivative of the heat capacity has a finite
discontinuity, which can indicate a third-order phase transition.
To be sure about that, we calculate the energy density for the
same samples of the particle-number densities, the functions
$\varepsilon(T)$ are shown in Fig.~\ref{fig:cv-ideal} in the right panel.
We see that these functions are continuous and smooth at  $T = T_{\rm c}$
and this proves an absence of latent-heat release at the critical temperature.

Let us briefly summarize the results obtained for an interacting
particle-antiparticle boson system, where the isospin (charge) density $n_I$
is conserved, and for a single-component ideal gas, where the particle-number
density $n$ remains constant.
First of all, we claim that
they both have the same critical curve $n_{\rm lim}^{\rm (id)}(T)$.
Furthermore, when $n^{(-)}(T)$, obtained for an interacting system,
and $n(T)$, obtained for an ideal gas, intersects the critical curve
$n_{\rm lim}^{\rm (id)}(T)$,
respectively, both systems undergo a phase transition of the second-order
or following the Ehrenfest classification of the third order.

It has long been known, see ref.~\cite{london-nature-1938},
that the Bose-Einstein condensation is indeed a third-order phase transition
according to the first classification of general types of transitions between
phases of matter, introduced by Paul Ehrenfest in 1933
\cite{ehrenfest-1933,jaeger-1998}.
Therefore, the obtained temperature $T_{\rm c}$ is really the temperature of the
phase transition of the second order (according to modern terminology) and the
density of condensate $n_{\rm cond} = n_{\rm cond}^{(-)}$ provided by $\pi^-$
mesons is the order parameter.

\section{Concluding remarks}
\label{sec:conclusions}

In this paper, we have presented a thermodynamically consistent method to
describe at finite temperatures a dense bosonic system that consists of
interacting particles and antiparticles at a fixed isospin density $n_I$.
We considered the system of meson particles with $m = m_\pi$ and zero spin,
which we named conventionally as ``pions''
because the charged $\pi$-mesons are the lightest nuclear particle and
the lightest hadrons that couple to the isospin chemical potential.

It turns out that the introduced dimensionless quantity $\kappa = A/2\sqrt{mB}$,
which is itself a combination of the mean-field parameters $A$, $B$ and the
value of a particle mass, is the scale parameter of the model.
Furthermore, it determines the different possible phase scenarios which
occur in the particle-antiparticle boson system.
Attraction coefficient $A = \kappa A_{\rm c}$, where
$A_{\rm c} \equiv 2\sqrt{mB}$, was parameterized by  $\kappa$ with
$\kappa = 1$ as the critical value that separates the regime of a
``weak attraction'' ($\kappa \le 1$) from the regime of a ``strong attraction''
($\kappa > 1$).
In this paper, we only looked at the ``weak attraction'' case.

It was shown that in the particle-antiparticle meson system, where the isospin
density $n_I$ is conserved, there is a Bose-Einstein condensate in the system in
the temperature interval $0 \le T \le T_{\rm c}$, which is the result of a
second-order phase transition that occurs at a temperature $T_{\rm c}$
and condensate density is an order parameter.\footnote{Note, the chiral
perturbation theory predicts that transition between the vacuum and the BEC
state is of the second order with universality class $O(2)$ \cite{son-2001}.}
This statement is in contrast to the conclusion given in
Refs.~\cite{anchishkin-mishustin-2019,anchishkin-4-2019,stashko-anchishkin-2020,
mishustin-anchishkin-2019},
where the system with zero chemical potential, $\mu = 0$, was investigated.
Indeed, in these works it was shown that in the case of a sufficiently strong
attractive mean field ($\kappa > 1$), the multibosonic system undergoes a
first-order phase transition and, as a result, develops a Bose condensate,
starting from a finite temperature.

So, we obtained that independently of parameters of the mean field, the
multi-boson system develops the Bose condensate for particles of the
high-density component only.
This means that in the pion gas, where $n_I = n_\pi^{(-)} - n_\pi^{(+)} > 0$,
the  $\pi^{-}$ mesons only  undergo the phase transition to the Bose-Einstein
condensate.
At the same time, the $\pi^{+}$ mesons exist only in the thermal phase for the
whole range of temperatures.
Then, for the experimental efforts, it makes sense to look for the Bose
condensate, which is created just by $\pi^{-}$ mesons.

For the description of the system's thermodynamic properties, we use the
Canonical Ensemble formulation, where the chemical potential $\mu$
is a thermodynamic quantity that depends on the canonical variables $(T,n_I)$.
We calculated dependence of the chemical potential on temperature for
different attraction parameters $\kappa$ which show that $\mu \approx$~const
in the ``condensate'' interval of temperatures $0 \le T \le T_{\rm c}$,
where these constant values  depend on the intensity of attraction.
Meanwhile, the temperature $T_{\rm c}^{(-)}$ of the phase transition to
the Bose-Einstein condensate of $\pi^-$ mesons (high-density
component) exhibits weak dependence on $\kappa$, as one can see in
Fig.~\ref{fig:2nd-order} in the left panel.
For all values $0 \le \kappa \le 1$ that we have considered these
critical temperatures differ from one another not more than $4$~MeV, this
inspires an introduction of
the mean value $T_{\rm c} = \left\langle T_{\rm c}^{(-)} \right\rangle $ of
the phase transition to the Bose-Einstein condensate.

The results obtained are in correspondence with known peculiar property
of the ideal Bose gas:
the Bose-Einstein condensation represents the third-order phase transition
or a discontinuity of the derivative of the specific heat
\cite{london-nature-1938}.
In the framework of the presented model, we obtained that in the same way,
the derivative of the specific heat undergoes a break at the temperature
$T_{\rm c}$, as one can see in the left panel in Fig.~\ref{fig:cv-s}.
The smooth dependencies of the energy density and entropy density on temperature
and the absence of latent heat release at $T_{\rm c}$ can be seen in the right
panel of Fig.~\ref{fig:cv-s} which proves that the system actually undergoes a
second order phase transition at this temperature.

The role of neutral pions is left beyond the scope of the present paper.
The present analysis can be improved by addressing these issues in more detail
and generalizing the calculation to nonzero contribution to the mean field that
depends on $n_I$.
The authors plan to consider these problems elsewhere.

\section*{Acknowledgements}
We thank H.~Stoecker for support and I.~Mishustin and V.~Vovchenko for reading
the manuscript and making valuable remarks.
The work of D.~A. is supported by the National Academy of Sciences of Ukraine
by its priority project "Fundamental properties of the matter in the relativistic
collisions of nuclei and the early Universe" (No. 0120U100935).
The work of  D.~Zh. is supported by Program "The structure and dynamics of
statistical and quantum-field systems" of the Department of Physics and Astronomy
of NAS of Ukraine.


\end{document}